# Peak into the Past
# An Archaeo-Astronomy Summer School

Dr. Daniel Brown, Natasha Neale, and Robert Francis

**Description:**
The delivery and impact of an archaeo-astronomy summer school for year 5-8 pupils, utilising local Peak District National Park monuments as an outdoor classroom.

**Abstract:**
Our landscape has been shaped by man throughout the millennia. It still contains many clues to how it was used in the past giving us insights into ancient cultures and their everyday life. Our summer school uses archaeology and astronomy as a focus for effective out-of-classroom learning experiences. It demonstrates how a field trip can be used to its full potential by utilising ancient monuments as outdoor classrooms. This article shows how such a summer school can be embedded into the secondary curriculum; giving advice, example activities, locations to visit, and outlines the impact this work has had.

**Key Words:** Astronomy, Archaeology, Enrichment, Outdoor classroom

**How it all began**
Over recent years, initiatives have been brought forward to encourage learning outside the traditional classroom environment, especially in topics such as physics and mathematics. These new approaches are used to illustrate how Science, Technology, Engineering, and Mathematics (STEM subjects) relate to everyday life. Traditionally, locations associated with physics and mathematics based activities were planetariums and science parks. National Parks and ancient monuments were related to ecological, biological, and geographic field work (Braund & Reiss, 2005).

Given the recent interest in astronomy during the International Year of Astronomy (2009), the five day summer school focused on the topic of archaeo-astronomy. This topic acts as an ideal melting pot for many subjects covered in school: astronomy, physics, geography, biology, religious studies, ICT and archaeology. Many subjects are STEM related and include aspects of citizenship.

The summer school took place in 2009 and was carried out together with The Meden School and Technology College, Worksop, to support student transition from primary to secondary schools. 27 Gifted and Talented (G&T) students participated, covering age groups from years 5 and 6 at primary and 7 and 8 at secondary school.

**What needed to be planned?**
The initial planning of the summer school activities was supported by working closely together with a core group of four G&T students in the months before the actual event. They were given the topic of 'What is archaeo-astronomy?' to research. Their results were used to assess the amount of knowledge and initial misconceptions related to archaeology and astronomy. This allowed us to develop activities that were more suited to the age group.

From their feedback, it soon became clear that the larger and unique ancient monuments (e.g. Stonehenge) were well known by the parents and the teachers. However, local and equally impressive monuments were unknown. This led to the exploration of sites that were only an hour's drive away from the school, as outdoor classroom environments.

Additionally, the movement of the rising and setting locations of the Sun during the seasons were not well known or merely generalised to East and West. This is a common misconception and a result of our modern lifestyle with light pollution and built up areas obstructing the views of the horizon. Therefore, the determination of the alignment of some sites was intended. The results would then allow the students to compare the orientation of the sites with the varying rising and setting points of the Sun during the seasons. This involved the introduction of Stellarium (a simple planetarium software, see Stellarium 2010) and the use of a magnetic compass. It also allowed the students to understand how able ancient mankind was with regard to construction and observations.

Two local sites were chosen, around which the summer school was developed: Arbor Low and Nine Ladies (see Box 1 for more details). The United Kingdom has an abundance of such sites and other examples can be located using Burl (1995).

The size and structure of Arbor Low lent itself to develop activities that incorporated area and volume calculations. These could then be used to determine the weight of stones and estimate the time it took to build this site. Also, the site shows clear evidence of its changed use during the millennia and signs of the earliest excavations by the first archaeologists (antiquarians) in the 18$^{th}$ century (see McGuire & Smith, 2008).

Its location within a field used by sheep was in stark contrast to the site of Nine Ladies that is in the middle of Stanton Moor. Ecological comparisons were made between the two sites using methods such as quadrat and transect sampling, soil pH, exposure, and temperature. A pre-visit was required to establish the initial use of the land and understand the land usage during the past, to develop appropriate activities supporting such work. Additional knowledge of the flora of the area was required to assist the pupils with the identification of species during sampling.

Pre-visits were undertaken for both sites to additionally assess any risks and help to comply with out-of-school activity guidelines. This also included establishing a meeting point in case of bad weather and determining the required (specialist) supervision ratio. More detail on the logistics around Nine Ladies at Stanton in Peak can be found in Box 2.

**Making it all come together**
The summer school itself consisted of five days. During the first two days activities based at the Meden School were carried out. These were designed to introduce the topics covered in the summer school and develop the required skills. These days incorporated some teacher focused sessions but were overall student focused, using many practical experiments and competitions. Day three was dedicated to visiting the two ancient monuments. The collected data was then analysed on day four during a workshop at the Centre of Effective Learning in Science (CELS) at Nottingham Trent University (NTU). During this time the group also visited the Observatory at NTU and experienced a planetarium session.

Day five was a presentation day at which the students presented the results of their work to their parents at a specifically booked conference venue within the working day. They planned the event, designed the displays and invited guests. Furthermore, they prepared the required props for their presentation. During this phase they were in total control of the creative process which allowed the assessment of the impact summer school had on the entire group.

**BOX 1 Interesting Peak District monuments**

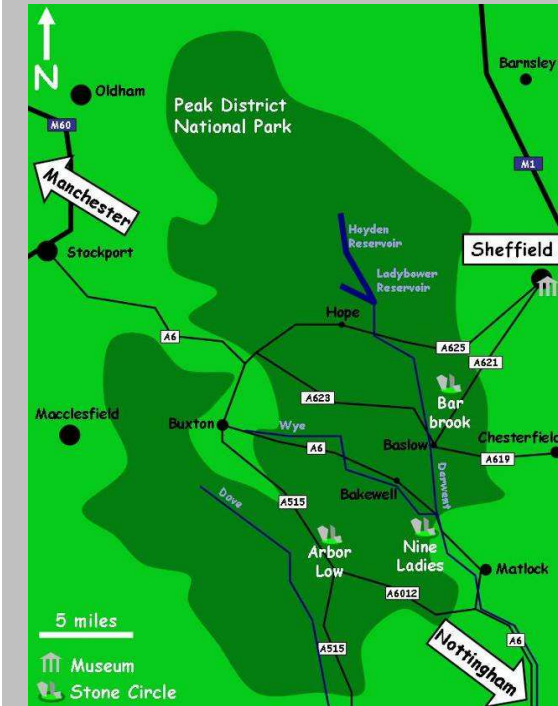

**Figure 1** A map of the Peak District National Park indicating its general location and points of interest.

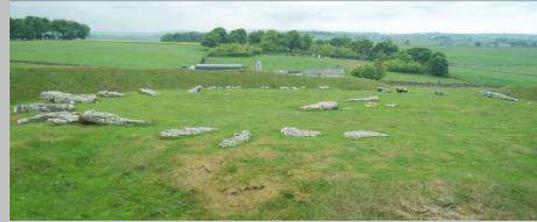

**Figure 2** A panoramic view of the central part of Arbor Low.

**Arbor Low**

A large stone circle constructed in multiple phases and consisting of limestone stones (some 4 m tall) now fallen over and surrounded by a large ditch and embankment (see figure 2). The monument is nearly 100 m in diameter and lies within a meadow. It shows clear evidence of former archaeological digs by the antiquarian T. Bateman.

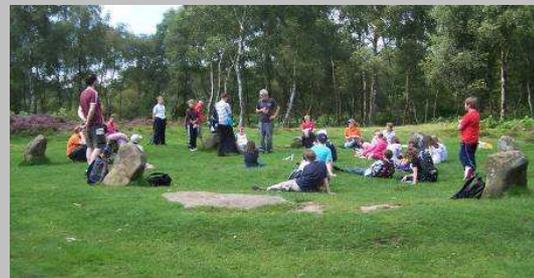

**Figure 3** The summer school group at Nine Ladies.

**Nine Ladies**

A smaller stone circle made of gritstones and only 10 m in diameter (see figure 3). It is located in the burial area on Stanton Moor. It shows some signs of a Victorian wall surrounding it and is linked to recent protests against quarrying close by.

The Peak District National Park is Britain's first national park and is located as shown in figure 1 between the two metropolitan areas of Manchester and Sheffield. Its landscape has been changed by mankind from the early Bronze Age up until now. For the summer school two sites were selected that not only offered impressive stone monuments, but also were situated in different landscapes.

When visiting any sites on non-public access land, care should be taken to inform the landowner beforehand. Arbor Low is such an example. Contact B. Walley of Mosar Farm, Ashbourne Road, Monyash.

**BOX 2 Nine Ladies**

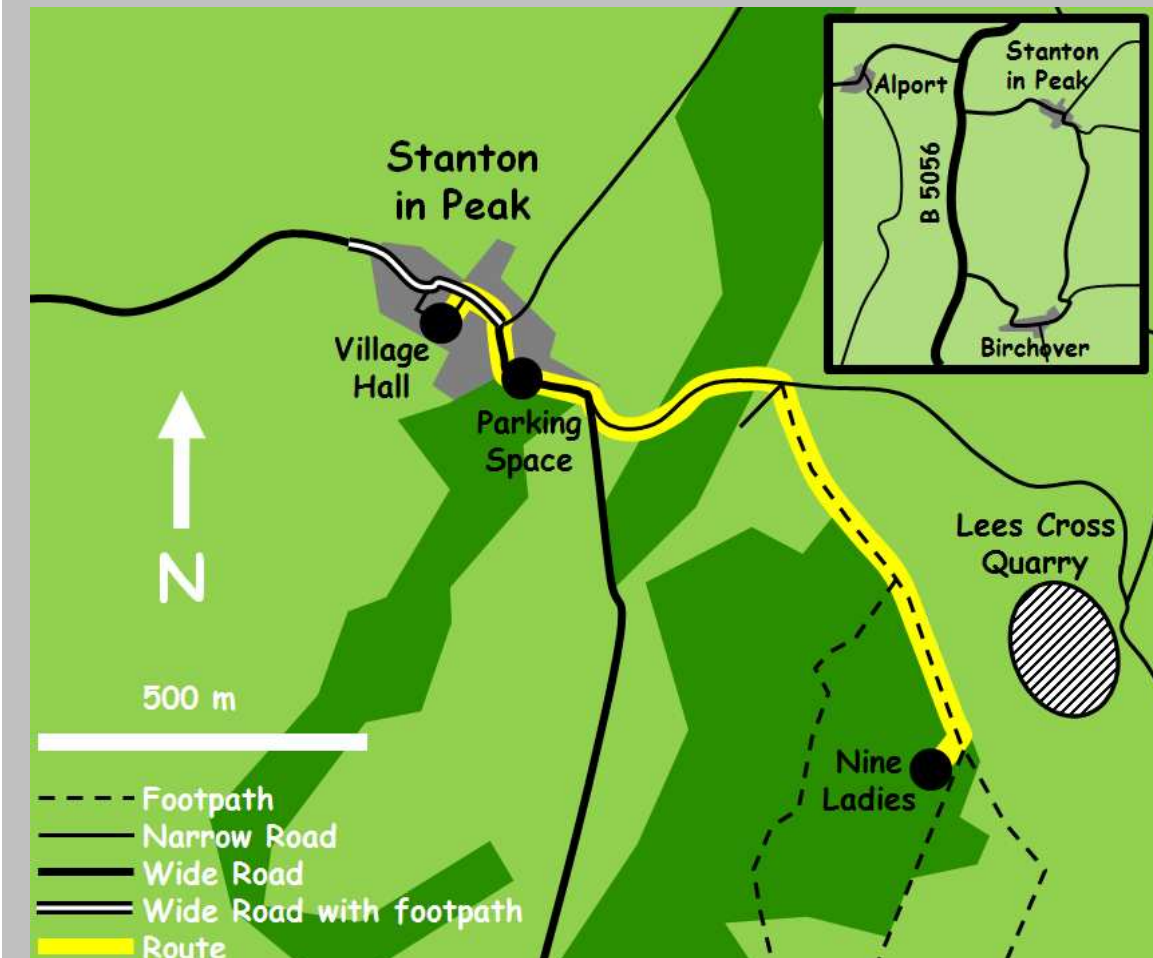

**Figure 4** A map of the Stanton in Peaks area and a possible route to access Nine Ladies.

The site of the Nine Ladies can be reached via Stanton in Peak. A bus can park in a wider area along the village street indicated in figure 4. It is not easy to turn in the village and it is advisable to take a route through Birchover to rejoin the B5056 again (see inset). The group can walk to the village hall that provides toilet facilities and a large open space both inside and outside to relax and have lunch. Bookings must be made prior to the trip. There are no eating facilities in the village except for a small pub. The actual site is only a short 1.5 km walk away. Some of the walk is along quiet roads that in parts do not have dedicated pavements. However, half of the walk is over public footpaths off road that are uneven and rocky in places, and can be muddy. Therefore sturdy footwear is advised. Keep following the main footpath and you arrive at Nine Ladies to your right hand side. Stanton in Peak and the Stanton Moor region are frequently visited by tourists. However, if other sites are visited, care should be taken to ensure the support of the local community.

**Stanton in Peak – Village Hall**
At the time of writing, bookings can be made via Mrs. Corran for a charge of £4 + VAT per hour for non-residents.

Address:        Middlestreet,
                Stanton in Peak
Telephone:      01629 636915

To illustrate some of the activities and how they make use of the outdoor classroom environment, an example of activities will briefly described linked to:
"How much work was needed to construct an ancient monument?"

- **Stone Race**
  Move a stone a given distance.
  Best method of transporting a stone determined by the duration of travel
  Methods tested: pulling with ropes (figure 5), dragging, or pushing.
  The modelled stone consisted of a plastic box filled with sand.
  Results are scaled up to realistic mass and distances.

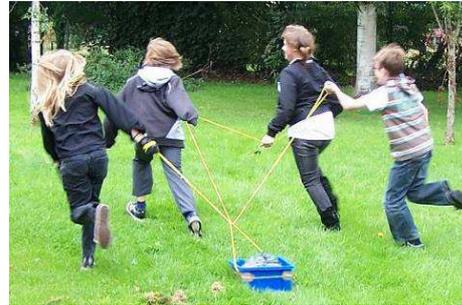

**Figure 5** A team using the pulling approach for the Stone Race in the school grounds.

- **Dig a Hole**
  Dig a hole of given dimensions.
  Best method to dig ditches determined by the duration of the digging.
  Dig a hole of fixed dimensions (i.e. given volume) in a sandpit.
  Methods tested: bare hands, replica cow scapula (shoulder blade), or small plastic spades (figure 6)
  Results are scaled up to realistic volumes.

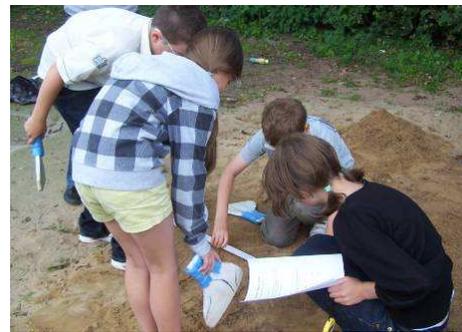

**Figure 6** A team digging a hole with replica of a scapula in the school grounds.

- **Surveying Arbor Low**
  Survey dimensions of site (selected stones as in figure 7 and ditch).
  Ditch was assumed to be circular and its cross-section a triangle.
  Discuss the simplifications and results: ditch was much deeper in the past.

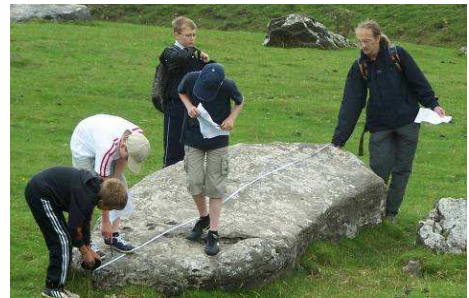

**Figure 7** A group measuring the volume of a central stone on site at Arbor Low.

- **Density Workshop**
  Determine the density of limestone (similar to Arbor Low).
  Volume and mass were measured using displacement vessels (figure 8) and scales.
  
  Calculation of $\text{density} = \dfrac{\text{mass}}{\text{volume}}$
  
  Apply result to surveyed stones at Arbor Low to determine their mass.

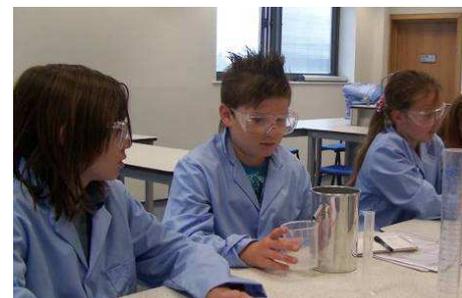

**Figure 8** A group using a displacement vessel to determine the volume of a limestone in the CELS science laboratories.

**BOX 3 Archaeobotany Workshop**

This workshop allowed students to reconstruct the changing landscapes during the transition from Neolithic hunter gatherers to semi-sedentary communities and on to the first farmers. Students built up a picture of the ancient landscape by analysing charred seed assemblages; remains of which are often found on archaeological sites. (Care should be taken not to ingest plant remains.) A simplified visual identification key was provided in figure 9, along with a basic environmental categorisation table (Table 1) which allowed students to associate certain species of plant with key types of environments (forests, meadow, and farmland). The transition from hunter gatherer to farmer is one of the pivotal moments in the prehistory of Britain, having long lasting effects on the culture and environment of the country.

Hazelnuts come from a tree (8 m high) often found in wood land or hedgerows. The nuts can be eaten and the wood used in construction.

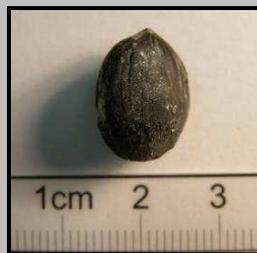

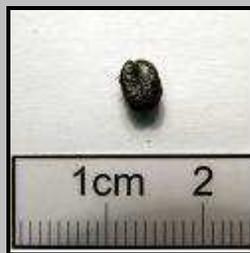

Wheat is a tall grass like agricultural crop planted in fields annually as a cereal. The plant is no native to Britain but was introduced in the Neolithic period.

Hawthorn berries come from a tree (4 m high). They are often found on hedgerow around fields where they are planted to form a

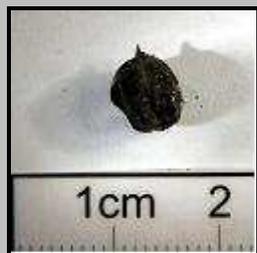

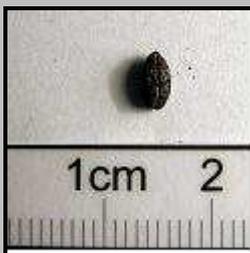

Oats are a tall grass like agricultural crop similar to wheat. The plant is grown annually. The plant is no native to Britain but was introduced in the Neolithic period.

Grass is found growing in open spaces is little or no shade. Wild and domestic animals eat grass.

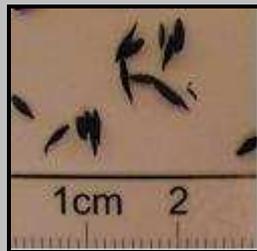

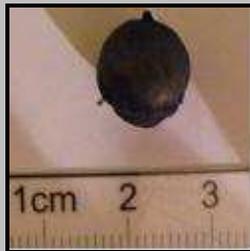

Oak acorns come from a tree (30 m high) found in forests. This tree can live for 900 years. The wood is excellent for construction.

**Figure 9** An example of a visual identification key.

**Table 1** The Environmental Categorisation Key

| Environmental Categorisation Key | | | |
|---|---|---|---|
|  | **Forest** | **Meadow** | **Farmland** |
| **Hazelnut** | ● | ● | ● |
| **Wheat** |  |  | ● |
| **Hawthorn** |  | ● | ● |
| **Oats** |  |  | ● |
| **Grass** |  | ● |  |
| **Oak Acorn** | ● |  |  |

**Creating a reference collection**

Plant remains need to be dry; collected at the point of harvest and wrapped tightly in tin foil. Wrap wood sections individually and seed/nuts in small quantities. Place into a laboratory furnace or kitchen cooker at 220-240°C for four hours. Unwrap cooled packets wearing laboratory gloves.

These activities illustrate how basic numeracy can be used to scale up results and apply them to tangible monuments. The outdoor classroom and ancient monument is transformed into a mathematics laboratory. The experiments carried out at school illustrate how to simplify problems and test initial assumptions. Furthermore, applying their result in the outdoor classroom environment allowed the students to experience the simplifications of the models developed in the lab.

**How it all fits in**
**Archaeology:** A further aspect targeted was how mankind has changed his environment in the past (Barnatt & Smith, 2004). At the sites this could not only be seen in the multiple construction phases of the monument itself but also by the ecology on site. The history of the monuments was explored by introducing basic concepts of archaeology including relative dating and stratigraphy, as well as the now outdated but intuitive three ages of Stone, Bronze, and Iron Age. Due to the modern scientific nature of the profession it is easy to find areas that can be used within the confines of the National Curriculum in England. At KS3, stratigraphy can be used in the context of geological activity caused by chemical and physical process (3.4a) that is part of 'The environment, Earth, and the Universe' program of study. Furthermore at KS4 archaeobotany can be integrated into the 'Organisms and health' program by exploring how organisms are interdependent and adapt to their environment (2.1a). Additionally, it also fits into the 'Environment, Earth, and Universe' program when seen in the context of the effects of human activity on the environment (2.4a). This link is well illustrated in the archaeobotany workshop. More details can be found in Box 3.

**Ecology:** The ecology studies on site were used to hypothesise how the use of the land influenced the present flora. The students were able to investigate its use by sampling and identifying the species present at the site and linking their results to the soil pH. Quadrat and transect sampling were simple for the students to carry out and organise themselves with little support. The students concluded that the pasture land usage at Arbor Low caused (or allowed) fewer shrubs to develop, in contrast with the land at Nine Ladies which has not (perhaps never) been grazed and where e.g. gorse and trees were found. Students also became confident in making their own observations and placing their findings into the context of the higher number of visitors at Stanton Moor and Nine Ladies compared with Arbor Low. The different exposure to light and wind at both sites could also be linked by the students to the greater diversity of species at Arbor Low. Overall the activities fit into the National Curriculum in England at all Key Stages particularly under Sc2 Life processes and living things. From Key Stage 1 Sc2 5a 'Find out about the animals and plants that live in local habitats' through to Key Stage 4 Sc2 5b 'How the impact of humans on the environment depends on social and economic factors'. However the skills employed by the students in taking measurements and recording data apply across the age groups to the investigative skills section of the curriculum.

**Citizenship:** The Nine Ladies site is ideal to introduce the problems of quarrying in the Peak District as well as aspects of citizenship related to the program 'Rights and responsibilities' in KS 3 (1.2b) (see McGuire & Smith, 2007). The site has been the location of a protest about quarrying from 1999 – 2008 and was covered in the media (see Ward, 2005).

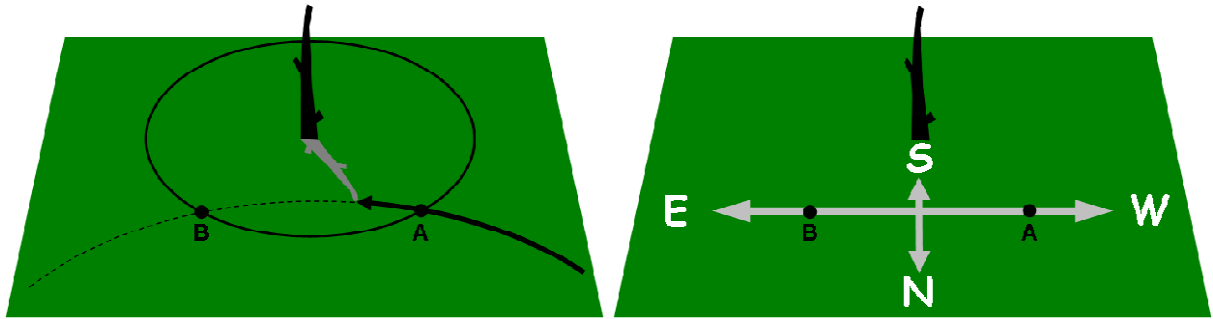

**Figure 10** Illustration of the Indian Circle method: First, draw a circle around an upright stick. Then wait until the tip of the shadow crosses the circle once in the morning (A) and again in the afternoon (B) as shown in the left hand side. Connecting both points allows you to determine the East-West and North-South axis demonstrated in the right hand side.

**Astronomy:** The topic of astronomy was covered by revisiting basic concepts related to shadows covered in 'Physical processes' topic at KS 2 (Sc4 3b), as well as seasons mentioned in 'The Earth and beyond'. Many activities can be found in Kibble (2010).

The students applied their knowledge to explain how Stone Age burials (Schmidt-Kaler & Schlosser, 1984) or other ancient monuments were aligned. During their work they were introduced to the prehistoric method of the 'Indian circle' in a workshop using a miniature sundial. This method is explained in figure 10 and allows the determination of the cardinal points. Furthermore this activity consolidates the material covered in 'Environment, Earth, and Universe' at KS 3 (3.4b), since it provides insight into the nature and observed motion of the Sun. The alignments from the centre of Arbor Low towards the openings in the ditch were checked to see if they relate to any seasonal positions of the Sun on the horizon. This activity was targeting the misconception of the constant rising and setting position of the Sun, since Arbor Low does not show any clear alignments.

**Did it Work?**

The impact of the project was judged on the student's choice and content of topics presented during the final day. Small groups of two to four students presented topics of their choosing that stood out for them in this summer school. As a result the students covered all the areas ecology, archaeology, and astronomy we had targeted. The content of these presentations illustrated that all predefined outcomes were achieved and important key words were included in their appropriate context, e.g. including the three age system into a theatre performance. Furthermore, the students presented their results using Power Point slides, photography, theatre performance, and poetry. This is an example of such a poem:

*Archaeoastronomy is what we learnt this week*
*Before we came to summer school about this subject we were bleak.*

*To start with the guys from CELS sent us out to seek*
*Some things about Stonehenge, so we thought we'd have a peak.*

*Next we did some experiments to find out how to roll*
*A heavy box of coco-pops using a large pole.*

*Then we did a big mock dig which meant going underground*
*And after all the layer and layers this is what we found.*

*We found some roman artefact, coins, swords, and bones.*
*There were also some old ancient tools – which were of course made of stone.*

*On Wednesday we went to two heritage sites both different in their way.*
*Measured ditches, stones, learnt about hills and bones – so were sleepy at the end of the day.*

> *It's Thursday, we're at University, looking through a big telescope.*
> *Got to prepare those presentations tomorrow – I don't know how we'll cope.*
>
> *Friday has come at last but not least, we're sad that we're leaving – Oh No!*
> *But after all our hard work and effort we hope you enjoyed the show.*

The use of these different approaches reflects skills linked to individual growth and improvements in social skills which could especially be observed with the initial focus group.

**A summer school legacy**
To further support the skills the students developed in the summer school, follow up work for more than a year was undertaken. This work included the set up of a web based environment (WIKI), development of Continuing Personal Development (CPD) activities, and a visit to Weston Park Museum, Sheffield.

The WIKI pages developed by the students (see Brown et al. 2010) allowed them to reflect upon their work and use both literacy and IT skills to communicate their findings. Their webpage has also remained a vital tool for keeping in contact with the students and monitoring their progress, given that the specialist scientists of the summer school are not based at the school. Additionally, it also allowed establishing international links that enabled students to discuss their questions with foreign scientists and make the students more confident.

A spin off from the summer school was the intensive use of Stellarium in learning and teaching. The initial focus group of four G&T students developed on their own a CPD activity and resources, both printed and IT based, to introduce teachers to Stellarium. The CPD activity was presented by them at the Annual ASE conference January 2010 in Nottingham. Their work demonstrates again how the summer school has increased their self confidence and allowed them to follow up on some aspects mentioned.

In March 2010, a smaller group of summer school participants explored the Weston Park Museum in Sheffield. This visit allowed the students to see some of the artefacts discovered at the sites they had visited the previous year. The students demonstrated a considerable amount of knowledge that they had retained from the summer school. During the visit the students could place the sites into their local history and learn more about antiquarians, archaeological excavation methods, and curating a museum. Again, the WIKI page was an essential tool to support the work of the teachers beforehand, without the need of a specialist scientist having to visit the school.

As a result, the archaeo-astronomy summer school has created a flourishing legacy supporting both teachers and students. The summer school also provided tangible proof of how easy it can be to incorporate an interdisciplinary topic such as archaeo-astronomy into the curriculum of a secondary school on many different levels. We have also provided and developed many resources that can be used to develop workshops and organise field trips to use the outdoor classroom to its full potential. The example of Arbor Low and Nine Ladies is only one possible location and other schools in other areas will have a wide range of other ancient sites and landscapes waiting to be explored.

We feel that such work should be fully supported by schools, headteachers and parents since it has not only furthered the teaching and learning experience of the participating students, but also allowed the involved teaching staff to overcome many barriers associated with the use of the outdoor classroom (Rickinson et al., 2004). Its success has enabled us to develop a more general project supporting the outdoor classroom.

**Position held:**

**Dr. Daniel Brown**　　is a professional astronomer, former astronomy lecturer, and member of the Initiativkreis Horizontastronomie im Ruhrgebiet e.V. promoting astronomy outreach in Germany. He is now Astronomy Outreach and Development Officer at Nottingham Trent University.
e-mail: daniel.brown02@ntu.ac.uk

**Natasha Neale**　　is a former conservation Biologist. She is now the Outreach Coordinator for the Centre of Effective Learning in Science at Nottingham Trent University.
e-mail: natasha.neale@ntu.ac.uk

**Robert Francis**　　is a former professional archaeobotanist with a PGCE. He is now a science technician at the School of Education in Nottingham Trent University.
e-mail: robert.francis@ntu.ac.uk

**References:**

Barnatt, J. & Smith, K. (2004) *The Peak District: Landscapes Through Time*, Windgate Press Ltd, Maccelsfield

Braud, M. & Reiss, M. (2005) *Learning science outside the classroom: bringing it all together*, Science Teacher Education, (**42**) 7

Brown, D. *et al.* (2010) *Archaeo-astronomy WIKI*, PBWORKS [online], Available at http://medenschool.pbworks.com [Accessed: 25 June 2010]

Burl, A. (1995) *A Guide to the Stone Circles of Britain, Ireland and Brittany*, Yale University Press, London

Kibble, B. (2010) *Sunshine, Shadows and Stone Circles*, Milgate House Publishers, Crewe

McGuire, S. & Smith, K. (2008) *Arbor Low and Gib Hill Conservation Plan* 2008, English Heritage

McGuire, S. & Smith, K. (2007) *Stanton Moor Conservation Plan* 2008, English Heritage

Rickinson, M., Dillon, J., Teamey, K., Morris, M., Young Choi, M., Danders, D., and Benefield, P. (2004) *A Review of Research on Outdoor Learning*, Field Studies Council [online], Available at http://www.field-studies-council.org/reports/nfer/index.aspx [Accessed: 25 June 2010]

Stellarium (2010), *Stellarium*, sourceforge.net [online] , Available at http://www.stellarium.org/ [Accessed: 25 June 2010]

Schmidt-Kaler, T. & Schlosser, W. (1984) *Stone-age Burials as a Hint to Prehistoric Astronomy*, R. A. S. Canada. Journal, (**78**) 5

Ward, D. (2005) *Eco-warriors sense victory in battle to protect Nine Ladies,* Guardian, 14 February, [online], Avaialable at
http://www.guardian.co.uk/environment/2005/feb/14/environment.society [Accessed 14 September 2010]